\documentclass[aps,prc,twocolumn,superscriptaddress,showpacs,showkeys]{revtex4}
\usepackage{graphicx}

\begin{document}


   \title{Effective operators from exact many-body renormalization}




\author{A. F. Lisetskiy}
\email[]{lisetsky@physics.arizona.edu}
\author{M.K.G. Kruse}
\author{B. R. Barrett}
\affiliation{Department of Physics, University of Arizona, Tucson, AZ 85721}

\author{P. Navratil}
\affiliation{ Lawrence Livermore National Laboratory,
Livermore, CA 94551}
\author{I. Stetcu}
\affiliation{Department of Physics, University of Washington, Box 351560, Seattle, Washington, 98195-1560}
\author{J. P. Vary}
\affiliation{Department of Physics and Astronomy, Iowa State University, Ames,
Iowa 50011}


\date{\today}

\begin{abstract}
We construct effective two-body Hamiltonians and E2 operators for the p-shell
by performing $16\hbar\Omega$ {\em ab initio} no-core shell model (NCSM)
calculations for A=5 and A=6 nuclei and explicitly projecting the
many-body Hamiltonians and E2 operator onto the $0\hbar\Omega$ space.  We then
separate the effective E2 operator into one-body and two-body contributions
employing the two-body valence cluster approximation. We analyze the convergence
of proton and neutron valence one-body contributions with increasing model space size and
explore the role of valence two-body contributions. We show that the constructed
effective E2 operator can be parametrized in terms of one-body effective charges
giving a good estimate of the NCSM result for heavier p-shell nuclei.
\end{abstract}

\pacs{21.10.Hw;21.60.Cs;23.20.Lv; 27.20.+n}
\keywords{NCSM, ab initio, effective interactions, E2 operator }

\maketitle

\section{Introduction}
In recent years, {\it ab initio} many-body nuclear structure calculations, such as the No Core Shell Model (NCSM) and Green's Function Monte Carlo (GFMC) have significantly progressed, to realistically describe heavier and heavier nuclei \cite{Nav07,Nog06,Ste05,Nav03,Nav00,Pie02,Pie01}. In the last few years, these calculations have been able to reproduce observables of light atomic nuclei up to A=14. To deal with heavier nuclei ($A \ge 15$), even at the current level of accessible computing power, it is unavoidable to adopt model space restrictions
which have to be accompanied by proper renormalization of bare NN and NNN interactions.
Significant efforts have been devoted to developing the coupled cluster theory with single and double
excitations (CCSD) \cite{Kow04}, the importance truncation scheme \cite{Rot07} and
to recast the {\em ab initio} NCSM approach by introducing a core and few-body valence
clusters \cite{Lis08}. Those studies are usually focused on binding energies and nuclear
excitation spectra. The electromagnetic and semi-leptonic operators, on the other hand,
 have been studied less frequently, and, less is known about their
 renormalization. One of the recent studies \cite{Ste05}, for example,
shows that the long-range quadrupole operator undergoes insufficiently weak renormalization
in the two-body cluster approximation.  This is in contrast to short-range strong interactions
and short-range operators, which are well-renormalized, even in the two-body cluster approximation.

 To explore the role of higher-body correlations for proper renormalization of the
long-range E2 operator, we can use the valence cluster expansion (VCE) considered in
Ref.\cite{Lis08}.
Since the effective p-shell interactions constructed in this approximation
account exactly for six-body correlations, it is also possible to construct the
effective two-body E2 operator which accounts for those six-body cluster correlations. \\
  In this paper, we present a detailed study of the properties of the effective E2 operator in the NCSM formalism when projected onto a single major shell.
The construction of our effective E2 operator, which acts in a $0\hbar\Omega$ valence space, is achieved as follows. We first performed a $N_{\rm max}\hbar\Omega$ NCSM calculation, for both $^5$Li and $^5$He, using the
non-local CD-Bonn potential \cite{Mac01}.
 This NCSM calculation uses as input the effective interaction for $^6$Li, obtained in the two-body cluster approximation. Here, $N_{\rm max}$ corresponds to the total oscillator quanta ($N$) above the minimum configuration and varies from 2 to 16. After a renormalization to the $N_{\rm max} = 0$ space, the resulting quadrupole moments and E2 matrix elements form the one-body part of the effective E2 operator for the $p$-shell. A $N_{\rm max}\hbar\Omega$ $^6$Li NCSM calculation is then performed for the same range of $N_{\rm max}$, as before. After a similar renormalization to the $N_{\rm max} = 0$ space, the matrix elements of the E2 operator in this case contain both one- and two-body parts. By using the results from the $^5$Li and $^5$He calculations, we are able to subtract the one-body contribution from the $^6$Li E2 matrix elements and are left with the pure two-body contribution. This step is necessary, as the effective operator will contain two-body contributions, even though the bare operator does not. We are, thus, able to construct an effective E2 operator from the one- and two-body
 contributions as a function of $N_{\rm max}$. We demonstrate that the two-body effective E2 operator can conveniently be parametrized in terms of
 j-dependent one-body effective charges. These effective charges account exactly for the one-body contributions as
well as averaged two-body contributions.

\section{Approach}

 \subsection{No Core Shell Model formalism}
 The starting point of the NCSM approach is the bare,
 exact  A-body Hamiltonian with the addition (and later subtraction)  of the Harmonic Oscillator (HO)
 potential \cite{Nav00}:

 \begin{equation}
 \label{hOmA}
 H^{\Omega}_{A} = \sum_{j=1}^{A}h^{\Omega}_j + \sum_{j>i=1}^{A}V_{ij}(\Omega,A),
 \end{equation}
  where $h^{\Omega}_j$ is the one-body HO Hamiltonian
  \begin{equation}
 h^{\Omega}_j = \frac{p^2_j}{2m} +\frac{1}{2}m\Omega^2r^2_j
 \end{equation}
 and $V_{ij}(\Omega,A)$ is a bare NN interaction $V^{\rm NN}_{ij}$, modified
 by the term introducing A- and $\Omega$-dependent corrections  to partly offset the HO potential present
in $h^{\Omega}_j$:
  \begin{equation}
  \label{vOmA}
  V_{ij}(\Omega,A)=V^{NN}_{ij}-\frac{m\Omega^2}{2A}(\vec{r}_i-\vec{r}_j)^2.
 \end{equation}
Another offset occurs for the center of mass (CM) part of the HO potential and the CM part of the kinetic energy operator, both present in Eq.(\ref{hOmA}), with the addition of a Lagrange constraint, $\Lambda_{CM} H_{CM}$, and with $\Lambda_{CM}$ chosen large and positive.  Here, $H_{CM}$ is the HO hamiltonian in the CM coordinates.
 The eigenvalue problem for the exact A-body Hamiltonian (\ref{hOmA}) for $A>3$ is
 very complicated technically, since an extremely large A-body HO basis is required to
 obtain converged results. The two-body cluster (2BC)  approximation,  consisting of
 solving Eq.(\ref{hOmA}) for the $a=2$-body subsystem of A particles, is commonly used \cite{Nav00}
 to construct effective two-body interactions for solving the A-body problem in restricted model spaces. However,
 the 2BC approach does not account for 3- and higher-body correlations in the effective interaction.
 To include the higher-body effects into the effective shell model interaction, we have recently developed \cite{Lis08} the two- and
 three-body valence cluster (2BVC and 3BVC) approximation, which includes higher-body correlations
 up to 6- and 7-body, respectively.
 We generalize this technique in order to compute effective shell model electromagnetic
 operators, as described below.

\subsection{Effective electromagnetic operators}
 Let us start with the assumption that we have derived a set of effective Hamiltonians ${\cal H}^{0, N_{\rm max}}_{A,a_1}$
for different values of $N_{\rm max}$ following the prescription given in \cite{Lis08}.
 According to the previously adopted notation, the first upper index (0) indicates that the effective Hamiltonian is
constructed for the $0\hbar\Omega$ space ( e.g., for the p-shell). The second upper index,
$N_{\rm max}$, indicates that the given effective Hamiltonian, when diagonalized in the p-shell, exactly
 reproduces the low-lying NCSM results obtained in the $N_{\rm max}\hbar\Omega$ space.
The second lower index, $a_1$, refers to the order of cluster expansion employed.
 Finally, the first lower index, A, specifies the mass number
of the nucleus, for which this effective interaction is constructed. For simplicity we will omit all these indices below and
will label the effective Hamiltonian, ${\cal H}_{J}$, corresponding to a subset of states with total
spin, J.
The eigenvectors of the effective Hamiltonian ${\cal H}_{J}$ form the unitary
transformation ${\cal U}_{J}$, which reduces ${\cal H}_{J}$ in the $0\hbar\Omega$ space to diagonal form:
\begin{equation}
\label{h-eigen}
E_{J} ={\cal U}_{J}{\cal H}_{J}{\cal U}^{\dagger}_{J}.
\end{equation}
This same eigenstate matrix ${\cal U}_{J}$ can also be used to calculate the matrix
elements of other effective operators, ${\cal O}_{A,a_1}^{\rm eff}(\lambda k;JJ')$,
between eigenstates with spins $J$ and $J'$ in the
$0\hbar\Omega$ space:
\begin{equation}
\label{op1}
 {\cal M}_{A,a_1}^{\rm eff}(\lambda k; JJ') =
 {\cal U}_{J} {\cal O}_{A,a_1}^{\rm eff}(\lambda k; JJ')
 {\cal U}^{\dagger}_{J'},
\end{equation}
where $k$ is the tensor rank of the
operator ${\cal O}_{A,a_1}(\lambda k; JJ')$; and $\lambda = E $ or $M$ denotes electric or
magnetic multipole radiation.
The required operator mapping procedure imposed on the matrix elements of
the effective operator ${\cal M}_{A,a_1}^{\rm eff}$, calculated with the eigenvectors in the
$0\hbar\Omega$ space, is that they are identical to the matrix
elements ${\cal M}_{A,1}^{\rm bare}$ of the bare one-body operator, obtained from the NCSM calculation
in the large $N_{\rm max}\hbar\Omega$ space, \emph{i.e.},
\begin{equation}
\label{me_id}
 {\cal M}_{A,a_1}^{\rm eff}(\lambda k;JJ') \equiv P_{a_1} {\cal M}_{A,1}^{\rm bare}(\lambda k;JJ') P_{a_1},
\end{equation}
where $P_{a_1}$ is a projector into the $a_1$-body 0$\hbar\Omega$ space.
A standard NCSM procedure exists for calculating effective operators, starting from the
bare operator in a large space \cite{Ste05}. That is, the matrix $M_{A,1}^{bare}$ is the original bare operator $O^{bare}_{A,1}$
transformed to the eigenbasis in the full $N_{max}$ space analogous to the one expressed in Eq.(5). This technique has been tested
 in a two-body cluster approximation;
however, it becomes cumbersome in the case of higher-cluster approximations. To overcome this problem, we first calculate
the bare matrix elements, ${\cal M}_{A,1}^{\rm bare}(\lambda k; JJ')$, using eigenstates obtained in the $N_{\rm max}$ space.
Then the matrix elements of effective operators can be determined
using the inverse transformation to the one given by Eq.(\ref{op1}), where the
 ${\cal M}_{A,a_1}^{\rm eff}(\lambda k; JJ')$ matrix is replaced by
the $P_{a_1}{\cal M}_{A,1}^{\rm bare}(\lambda k;JJ')P_{a_1}$
 matrix:
\begin{equation}
\label{op2}
  {\cal O}_{A,a_1}^{\rm eff}(\lambda k;JJ') =
 {\cal U}^{\dagger}_{J}P_{a_1} {\cal M}_{A,1}^{\rm bare}(\lambda k;JJ')P_{a_1}
 {\cal U}_{J'},
\end{equation}
 according to the definition ({\em i.e.}, Eq.(\ref{me_id})) of an effective operator.
The  effective ${\cal O}_{A,a_1}^{\rm eff}(\lambda k;JJ')$ operator can be represented in the standard shell
model (SSM) format, using the valence cluster expansion (VCE) (similar to the VCE for the effective Hamiltonian in \cite{Lis08}),
\begin{equation}
\label{opexp}
 {\cal O}_{A,a_1}^{\rm eff}=O^0_{A,A_c} + O^1_{A,A_c+1}+ \sum_{n=2}^{a_{\rm v}}O^n_{A,A_c+n},
\end{equation}
where the upper index, $n$, stands for $n$-body part of the effective operator in the $a_{\rm v}$-body
valence cluster ($a_1=A_c+a_{\rm v}$); the first lower index $A$, for the mass dependence; the second lower
index $A_c+n$, for the number of particles contributing to corresponding $n$-body part. The indices $JJ'$ and $\lambda k$ are
omitted on both sides of the equation for simplicity.


\section{Effective  $E2$ operator for the p-shell}

As an example we consider the $E2$ operator ({\em i.e.}, $\lambda =E$ and $k=2$) for nuclei in the p-shell
 with an s-shell core. Since the spin of the core is zero, there is no core contribution for $k=2$.
So the first term in the VCE, given by Eq.(\ref{opexp}), vanishes.
Taking $A=6$, we find, that the two-body VCE approximation
({\em i.e.}, when $a_1=A=6,a_{\rm v} =2$) is exact:
 \begin{equation}
\label{hncsm}
{\cal O}^{\rm eff}_{A=6,a_1=6}(E2) = O^{1}_{6,5}(E2) + O^{2}_{6,6}(E2),
\end{equation}
where the one-body part, {\em i.e.}, $O^{1}_{6,5}(E2)$,
is determined in terms of the proton and neutron one-body matrix elements for the p-shell
from the NCSM calculations for $^5$Li and $^5$He, respectively. To emphasize that $O^{1}_{6,5}(E2)$ is an one-body
operator in the p-shell space, we can represent it in the second-quantization formalism, {\em i.e.},
 \begin{equation}
\label{opv1sq1b}
 O^{1}_{6,5}(E2)=\sum_{ij}\langle i |O^{1}_{6,5}(E2) | j \rangle a^\dagger_i a_j,
\end{equation}
where the summation runs over all single-particle states considered ({\em i.e.}, $p_{1/2}$ and $p_{3/2}$),
$\langle i | O^{1}_{6,5}(E2) | j \rangle$ is the  matrix element for these states and
 $a^\dagger_i$ ($a_j$) is a single-particle creation (annihilation) operator.
Then the two-body part of the effective operator is calculated by
 \begin{equation}
\label{opv1}
 O^{2}_{6,6}(E2)={\cal O}^{\rm eff}_{6,6}(E2)-O^{1}_{6,5}(E2).
\end{equation}

\begin{figure}[h]
\includegraphics[scale = 0.20]{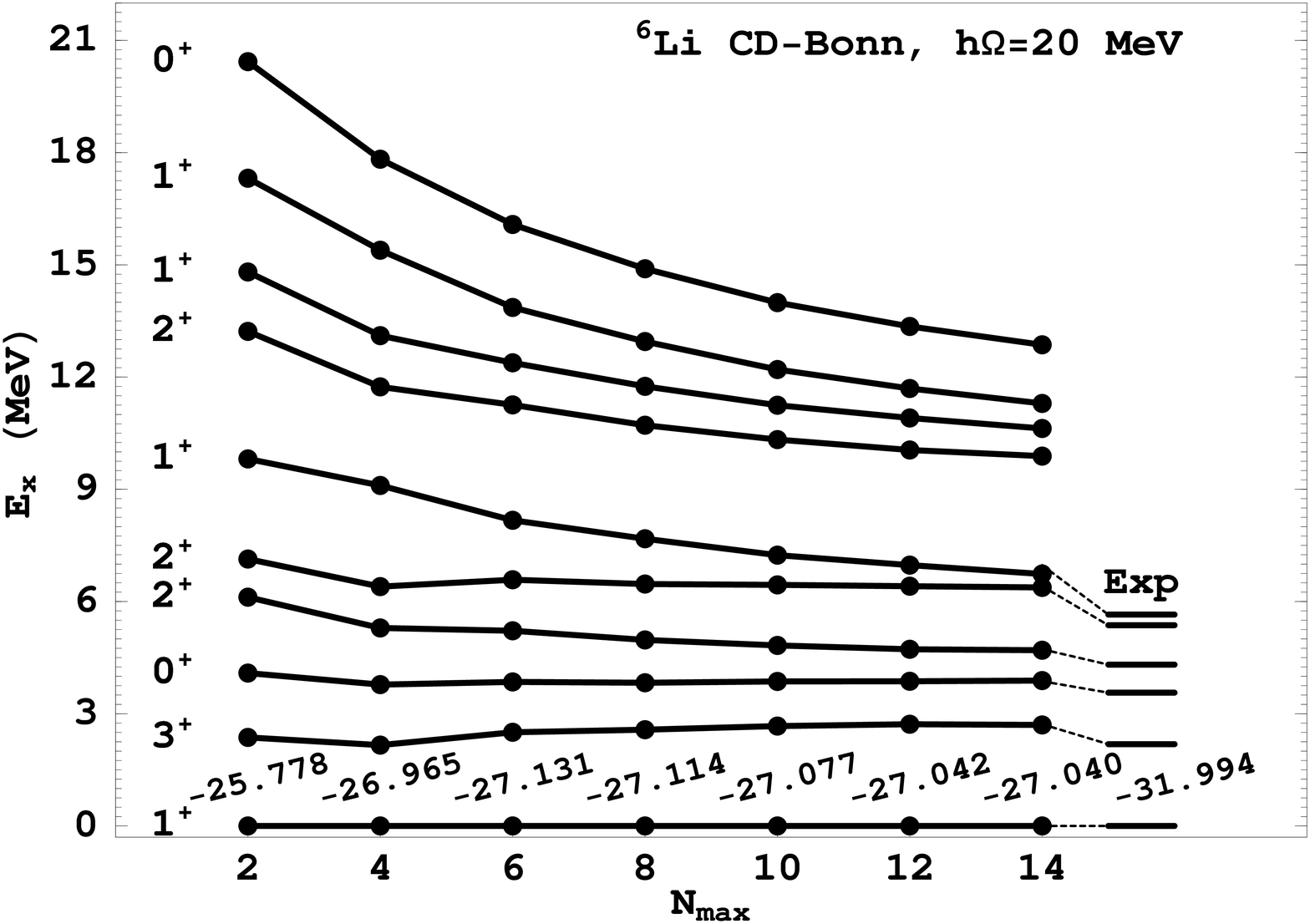}%
\caption{\label{spectra6li20hwcdb} NCSM results for the excitation energies of the $J^\pi$ states and the ground state energy
of $^{6}$Li, calculated in  $N_{\rm max}\hbar \Omega$ spaces with the CD-Bonn potential and $\hbar \Omega=20$ MeV.
The experimental spectra and the ground-state energy are shown for comparison.}
 \end{figure}

Again, to emphasize its two-body nature, we represent it in the second-quantization form:
 \begin{equation}
\label{opv1sq2b}
 O^{2}_{6,6}(E2)=\frac{1}{4}\sum_{ijsr}\langle ij | {\cal O}^{\rm eff}_{6,6}(E2) | sr \rangle a^\dagger_i a_j^\dagger
a_r a_s - O^{1}_{6,5}(E2),
\end{equation}
where the one-body operator, $O^{1}_{6,5}(E2)$, is given by Eq.(\ref{opv1sq1b}).
Utilizing the approach outlined in \cite{Lis08}, we have calculated the effective p-shell Hamiltonians for $^6$Li using the
6-body Hamiltonians with $N_{\rm max}=2,4,...,14$
and $\hbar \Omega= 20$ MeV constructed from the CD-Bonn potential \cite{Mac01}.
 To perform NCSM calculations we have used the specialized version of the shell-model code ANTOINE \cite{Cou99,Mar99}, adapted for the NCSM \cite{Cou01}.
The corresponding excitation
energies of p-shell dominated states and the binding energy of $^6$Li are shown in Fig.\ref{spectra6li20hwcdb}, as
 a function of $N_{\rm max}$.

Using the obtained wave functions for different values of $N_{\rm max}$, we have calculated the E2 matrix elements
${\cal M}_{JJ'}^{\rm bare}(E2)$. We have used the following relation,
$$b=\sqrt{\frac{\hbar}{M_{\rm p} \Omega}},$$
between the oscillator length parameter b (measured in fm) and harmonic oscillator frequency $\hbar \Omega$
(measured in MeV).
Then, using Eq.(\ref{op2}), we have determined the matrix
elements of the effective operator ${\cal O}^{\rm eff}_{6,6}(E2)$, which are listed in the last column of Table \ref{2body-e2}
 for $N_{\rm max}=14$.
 \begin{table}[tbp]
 \caption{\label{2body-e2} The values of reduced TBMEs of the one-body bare, one- and two-body
effective E2 operators for $^6$Li are shown in columns 7,8 and 9, respectively. The values of the
total reduced TBMEs, $D_t$ are given in the last column. These results correspond to $N_{\rm max}=14$.}
 \begin{tabular}{|cccc|cc|rr|rr|}
\hline
$2j_a$ & $2j_b$ & $2j_c$ & $2j_d$ & $J$ & $J'$ & \multicolumn{2}{|c|}{D$_1$, (efm$^2$)} & D$_2$, (efm$^2$) & D$_t$, (efm$^2$) \\
$\pi$  & $\nu$  & $\pi$  & $\nu$  &     &      &  bare & eff &  eff & eff \\
 \hline
 1 &   3 &   1 &   1 &   2 &   0 &         0.000 &        0.437 &        0.558 &        0.996 \\
 1 &   3 &   3 &   3 &   2 &   0 &         1.463 &        2.356 &        0.307 &        2.662 \\
 3 &   1 &   1 &   1 &   2 &   0 &        -2.070 &       -3.331 &       -0.537 &       -3.868 \\
 3 &   1 &   3 &   3 &   2 &   0 &         0.000 &       -0.309 &       -0.256 &       -0.565 \\
 3 &   3 &   1 &   1 &   2 &   0 &         0.000 &        0.000 &       -0.072 &       -0.072 \\
 3 &   3 &   3 &   3 &   2 &   0 &        -1.463 &       -2.254 &       -0.590 &       -2.844 \\
\hline
 1 &   1 &   1 &   1 &   1 &   1 &         0.000 &        0.000 &       -0.165 &       -0.165 \\
 1 &   1 &   1 &   3 &   1 &   1 &         0.000 &       -0.535 &       -0.825 &       -1.360 \\
 1 &   1 &   3 &   1 &   1 &   1 &         2.535 &        4.080 &        0.723 &        4.803 \\
 1 &   1 &   3 &   3 &   1 &   1 &         0.000 &        0.000 &        0.126 &        0.126 \\
 1 &   3 &   1 &   3 &   1 &   1 &         0.000 &       -0.311 &       -0.327 &       -0.638 \\
 1 &   3 &   3 &   1 &   1 &   1 &         0.000 &        0.000 &       -0.061 &       -0.061 \\
 1 &   3 &   3 &   3 &   1 &   1 &        -0.802 &       -1.290 &        0.101 &       -1.190 \\
 3 &   1 &   3 &   1 &   1 &   1 &        -1.792 &       -2.449 &       -0.278 &       -2.728 \\
 3 &   1 &   3 &   3 &   1 &   1 &         0.000 &        0.169 &        0.202 &        0.371 \\
 3 &   3 &   3 &   3 &   1 &   1 &         1.434 &        2.208 &        1.077 &        3.285 \\
\hline
 1 &   3 &   1 &   1 &   2 &   1 &         0.000 &        0.535 &        0.337 &        0.872 \\
 1 &   3 &   1 &   3 &   2 &   1 &         0.000 &       -0.311 &       -0.558 &       -0.869 \\
 1 &   3 &   3 &   1 &   2 &   1 &         0.000 &        0.000 &       -0.174 &       -0.174 \\
 1 &   3 &   3 &   3 &   2 &   1 &        -2.405 &       -3.870 &       -0.653 &       -4.523 \\
 3 &   1 &   1 &   1 &   2 &   1 &         2.535 &        4.080 &        0.372 &        4.452 \\
 3 &   1 &   1 &   3 &   2 &   1 &         0.000 &        0.000 &       -0.148 &       -0.148 \\
 3 &   1 &   3 &   1 &   2 &   1 &         1.792 &        2.449 &        0.761 &        3.210 \\
 3 &   1 &   3 &   3 &   2 &   1 &         0.000 &       -0.508 &       -0.146 &       -0.654 \\
 3 &   3 &   1 &   1 &   2 &   1 &         0.000 &        0.000 &        0.234 &        0.234 \\
 3 &   3 &   1 &   3 &   2 &   1 &        -1.792 &       -2.885 &       -0.332 &       -3.217 \\
 3 &   3 &   3 &   1 &   2 &   1 &         0.000 &       -0.379 &       -0.203 &       -0.582 \\
 3 &   3 &   3 &   3 &   2 &   1 &         1.603 &        1.913 &        0.395 &        2.308 \\
\hline
 3 &   3 &   1 &   1 &   3 &   1 &         0.000 &        0.000 &        0.145 &        0.145 \\
 3 &   3 &   1 &   3 &   3 &   1 &        -2.999 &       -4.827 &       -0.181 &       -5.009 \\
 3 &   3 &   3 &   1 &   3 &   1 &         0.000 &        0.634 &        0.584 &        1.218 \\
 3 &   3 &   3 &   3 &   3 &   1 &        -1.341 &       -2.066 &       -0.462 &       -2.528 \\
\hline
 1 &   3 &   1 &   3 &   2 &   2 &         0.000 &       -0.475 &       -0.643 &       -1.118 \\
 1 &   3 &   3 &   1 &   2 &   2 &         0.000 &        0.000 &       -0.039 &       -0.039 \\
 1 &   3 &   3 &   3 &   2 &   2 &         2.738 &        4.407 &        0.407 &        4.814 \\
 3 &   1 &   3 &   1 &   2 &   2 &        -2.738 &       -3.741 &       -0.731 &       -4.473 \\
 3 &   1 &   3 &   3 &   2 &   2 &         0.000 &       -0.578 &       -0.621 &       -1.199 \\
 3 &   3 &   3 &   3 &   2 &   2 &         0.000 &        0.000 &        0.177 &        0.177 \\
\hline
 3 &   3 &   1 &   3 &   3 &   2 &         2.449 &        3.942 &        0.424 &        4.366 \\
 3 &   3 &   3 &   1 &   3 &   2 &         0.000 &        0.517 &        0.425 &        0.943 \\
 3 &   3 &   3 &   3 &   3 &   2 &         2.449 &        2.921 &        0.185 &        3.107 \\
 3 &   3 &   3 &   3 &   3 &   3 &        -2.682 &       -4.131 &       -0.722 &       -4.853 \\
 \hline
\end{tabular}
\end{table}

To find the one-body part, $O^{1}_{6,5}(E2)$, of the $E2$ effective operator we have
performed similar NCSM calculations for $^5$Li and $^5$He (using the same interaction as for the $^6$Li calculation)
and have calculated one-body matrix elements of the effective E2 operator for protons and neutrons,
respectively. These are listed in Table \ref{1body-e2} for bare ($N_{\rm max}$=0) and effective
($N_{\rm max}$=12,14 and 16) E2 operator.Entries in italics represent extrapolated values as explained in the text.
 \begin{table*}[tbp]
 \caption{\label{1body-e2} The one-body reduced matrix elements of the effective one-body proton ($\rho=\pi$) and
neutron ($\rho=\nu$) E2 operators, $O^{1}_{6,5}(E2)$, using the CD-Bonn potential with $\hbar\Omega$ = 20 MeV
 (b=1.44 fm). The calculated one-body and secondary two-body effective charges are also shown.}
 \begin{tabular}{cc|cccccccccc}
 \hline\hline
 & & \multicolumn{10}{c}{$\langle j_a^\rho || O^1_{6,5}(E2) || j_b^\rho \rangle$, (efm$^2$)} \\
\hline
 & & \multicolumn{2}{c}{$N_{\rm max}$=0}  &  \multicolumn{2}{c}{$N_{\rm max}$=12}  &  \multicolumn{2}{c}{$N_{\rm max}$=14}  &  \multicolumn{2}{c}{$N_{\rm max}$=16} &  \multicolumn{2}{c}{$N_{\rm max} \rightarrow \infty$} \\
\hline
2j$_a$ & 2j$_b$ & $\pi$ & $\nu$ & $\pi$ & $\nu$ & $\pi$ & $\nu$ & $\pi$ & $\nu$ & $\pi$ & $\nu$ \\
 \hline
3      & 3      & -2.925  & 0.000 & -3.999  & -0.508 & -4.093  & -0.522 & -4.162   & -0.533 & {\em -4.524}   &  {\em -0.553} \\
1      & 3      &  2.925  & 0.000 &  4.711  &  0.618 &  4.847  &  0.636 &  4.933    & 0.650 &  {\em  5.417}   &  {\em -0.664} \\
\hline
 & & \multicolumn{10}{c}{$e_1^\rho(2j_a,2j_b)$} \\
\hline
3     & 3      &  1.000   & 0.000 & 1.367 & 0.174 & 1.399 & 0.179 & 1.422 & 0.183 & {\em  1.547} &  {\em 0.189} \\
1     & 3      &  1.000   & 0.000 & 1.610 & 0.211 & 1.656 & 0.217 & 1.693 & 0.222 &  {\em 1.852} &  {\em 0.227} \\
\hline
 & & \multicolumn{10}{c}{$e_2^\rho(2j_a,2j_b)$} \\
\hline
3     & 3      &  0.000   & 0.000 & 0.256 & 0.169 & 0.291 & 0.161 & {\em 0.333} & {\em 0.176}  & {\em  0.632} &  {\em 0.211} \\
1     & 3      &  0.000   & 0.000 & 0.182 & 0.209 & 0.221 & 0.244 & {\em 0.255} & {\em 0.248}  &  {\em 0.567} &  {\em 0.294} \\
\hline
 & & \multicolumn{10}{c}{$e_t^\rho(2j_a,2j_b)$} \\
\hline
3     & 3      &  1.000   & 0.000 & 1.623 & 0.343 & 1.690 & 0.341 & {\em 1.755} & {\em 0.359}  &  {\em 2.179} &  {\em 0.400} \\
1     & 3      &  1.000   & 0.000 & 1.792 & 0.420 & 1.877 & 0.461 & {\em 1.948} & {\em 0.469}  &  {\em 2.419} &  {\em 0.521} \\
\hline
\end{tabular}
\end{table*}

Using these one-body matrix elements of the effective one-body E2 operator, $O^{1}_{6,5}(E2)$,  we can determine the many-body matrix elements of that operator. Thus, in the considered case of two valence nucleons, one can calculate two-body reduced matrix elements (TBMEs) of the
one-body E2 operator using the following formula \cite{Bruss77}:
\begin{eqnarray*}
\frac{\langle j_a^\pi j_b^\nu; J || O^1_{6,5}(E2) || j_c^\pi j_d^\nu; J' \rangle}{\sqrt{(2J+1)(2J'+1)}}=
\end{eqnarray*}
\begin{eqnarray}
\label{2bodyME}
\left\{
\begin{array}{rrr}
  j_a & j_c &  2    \\
  J' & J & j_d  \\
  \end{array}
\right\}
 \langle j_a^\pi || O^1_{6,5}(E2) || j_c^\pi \rangle \delta_{b,d}f_{abJ'} +
\end{eqnarray}
\begin{eqnarray*}
\left\{
\begin{array}{rrr}
  j_b & j_d &  2    \\
  J' & J & j_a  \\
  \end{array} \right\}
\langle j_b^\nu || O^1_{6,5}(E2) || j_d^\nu \rangle \delta_{a,c}f_{cdJ},
\end{eqnarray*}
where $f_{abJ}=(-1)^{j_a+j_b+J}$ and the one-body matrix elements, $\langle j_a^\rho || O^1_{6,5}(E2) || j_c^\rho \rangle$,
are listed in Table \ref{1body-e2}. Using Eq.(\ref{2bodyME}), we have calculated the TBMEs of the one-body
bare ($N_{\rm max}=0$) and effective (for $N_{\rm max}=14$) E2 operators
$$D_1=\langle j_a^\pi j_b^\nu; J || O^1_{6,5}(E2) || j_c^\pi j_d^\nu; J' \rangle$$ and have listed them in
 columns 7 and 8 of Table \ref{2body-e2}, respectively.  Taking into account Eq.(\ref{hncsm}), we
calculate the TBMEs of the effective two-body E2 operator
$$D_2=\langle j_a^\pi j_b^\nu; J || O^2_{6,6}(E2) || j_c^\pi j_d^\nu; J' \rangle,$$
which are listed in column 9 of Table \ref{2body-e2}. Note, that by definition, $D_1$ and $D_2$ add to yield the
total TBMEs, $D_t = \langle j_a^\pi j_b^\nu; J ||{\cal O}^{\rm eff}_{6,6}(E2) || j_c^\pi j_d^\nu; J' \rangle$,
which we have determined using Eq.(\ref{op2}).

Analyzing the results shown in Table \ref{2body-e2}, we identify three types of the TBMEs:
\begin{enumerate}
 \item The TBMEs which have non-zero values for the bare one-body part, {\em i.e.,} $D_1$(bare)$\neq 0$:
       because in the case of the bare operator we have only the proton part contributing, this means that those
       TBMEs are allowed according to the one-body selection rules. The corresponding TBMEs of
       the  effective one-body part contain proton as well as neutron contributions, if the latter is
       not forbidden by the one-body selection rule for neutron single-particle orbitals. The TBMEs of the
       two-body part of the effective E2 operator, $D_2$, always contributes constructively to the total
       TBMEs and is about 20$\%$, on average, of the effective one-body part, $D_1$, in magnitude.
 \item The TBMEs which have zero values for the bare one-body part,  {\em i.e.,} $D_1$(bare) $= 0$, but
       non-zero values for the effective one-body part,  {\em i.e.,} $D_1$(eff)$ \neq 0$: in this case we have
       transitions which are forbidden according to the one-body selection rule for the proton and, subsequently,
       only the neutron contributes to the effective one-body part. The corresponding two-body parts are of
       the same order as the effective-one-body part and, as in the previous case, contribute constructively
       to the total TBME.
 \item  The TBMEs which have zero values for both bare one-body part,  {\em i.e.,} $D_1$(bare)$ = 0$,
        and effective one-body part,  {\em i.e.,} $D_1$(eff)$ = 0$: this means that such transitions are
        forbidden according to the selection rules for both the protons and the neutrons. Thus, there is only
        a non-zero two-body part, which is a factor of 2-3 less than the two-body part for the allowed
        transitions and is about an order of magnitude smaller than the one-body part for the
        allowed transitions of type 1.
\end{enumerate}

\subsection{Effective quadrupole charges}
To quantify the scale of the renormalization of the E2 operator, it is convenient to use
the traditional concept of effective quadrupole charges. The effective charges are
defined as rescaling parameters, which indicate how strongly the bare E2 operator has
to be enhanced in order to reproduce E2 matrix elements calculated in the
large $N_{\rm max}\hbar\Omega$ model space.

 Thus, it is helpful to define the one-body effective charges for specified values of
$N_{\rm max}$:
\begin{equation}
\label{cha1}
e_1^\rho(2j_a,2j_b) = \frac{\langle  j_a^\rho ||{\cal O}^{\rm eff}_{6,5}(E2) || j_b^\rho \rangle}
{\langle  j_a^\pi ||{\cal O}^{\rm bare}_{6,5}(E2) || j_b^\pi \rangle},
\end{equation}
where $\rho=\pi$ or $\nu$ denotes a proton or neutron, respectively. As one may note from
Table \ref{1body-e2} and Fig. (2),
 these one-body effective charges are $j$-dependent and have different values
 for protons and neutrons. There is stronger renormalization for protons than for neutrons, as
measured in the magnitude of the shift from their bare values (1 and 0, respectively). We also
note that the renormalization for the nondiagonal matrix element is somewhat stronger than for the diagonal one.
\begin{figure}[t]
	\centering
		\includegraphics[scale=0.33,angle=-90]{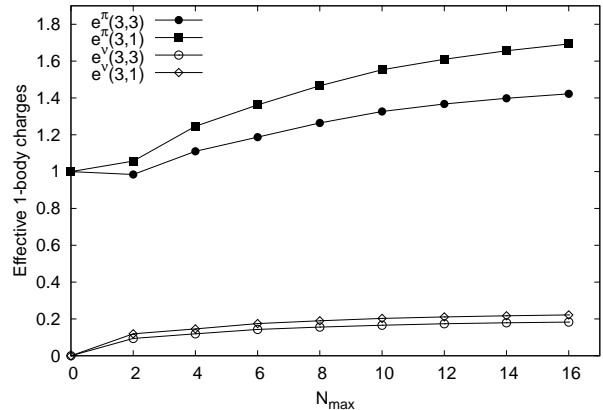}
	\caption{The one-body effective charges, $e_1^\rho(2j_a,2j_b)$, for $p$-shell space are shown
 as a function of increasing model space $N_{\rm max}$. }
	\label{fig:eff_charges_1body}
\end{figure}

To estimate the converged values of the effective charges,
$e_{\rm eff}(N_{\rm max} \rightarrow \infty)$, we fit an exponential plus
constant (see, {\em e.g.}, Ref. \cite{For08}) to each set of results for individual charges as a function of $N_{\rm max}$,
\begin{equation}
\label{fit}
e_{\rm eff}(N_{\rm max}) = a \exp{(-cN_{\rm max})} + e_{\rm eff}(N_{\rm max} \rightarrow \infty).
\end{equation}
The uncertainty in the extrapolating functional, assumed here as having an
exponential
form, should not produce errors larger than 10 $\%$.
 We have included the results for
$N_{\rm max}$ values from 2 to 14 in the fit.
 The extrapolated converged values are shown in the last two columns of Table
 \ref{1body-e2}.
 From the Table \ref{1body-e2}, we observe that the average magnitude of the proton
one-body effective charges for $N_{\rm max} =16$ appears similar to the standard phenomenological
value of 1.5. However, the average neutron effective charge is around 0.2 which is somewhat smaller
than the usual phenomenological value of 0.5. The estimated values of converged proton effective
charges are only about 10$\%$ larger than the corresponding $N_{\rm max} =16$ values, while the estimated
neutron converged effective charges are almost identical to the ones for $N_{\rm max} =16$,
which were not included in the exponential fit.

For the TBMEs of the effective E2 operator of type 1 and 2, one may model the two body part,
$D_2$, in terms of one-body matrix elements using relations similar to Eq.(\ref{2bodyME}).
\begin{figure}[b]
	\centering
		\includegraphics[scale=0.33,angle=-90]{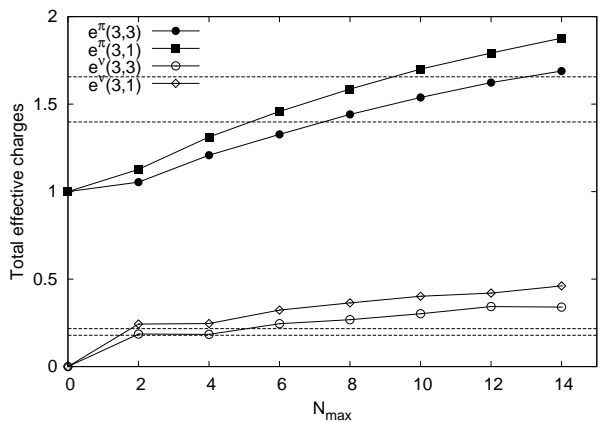}
	\caption{The total effective charges
$e_t^\rho(2j_a,2j_b)=e_1^\rho(2j_a,2j_b)+e_2^\rho(2j_a,2j_b)$ are shown as
 a function of $N_{\rm max}$. The dashed lines indicate the corresponding
  values of the one-body charges at $N_{\rm max} = 14$
  (see figure \ref{fig:eff_charges_1body}).}
	\label{fig:eff_charges_total_new}
\end{figure}
In the most general case we would have an additional one-body part depending not only on the
quantum numbers of the single-particle orbitals involved, but also on the spin values of initial
and final states, $J$ and $J'$. Alternately, we may introduce secondary ($J$ and $J'$ independent)
effective charges $e^\rho_2(2j_a,2j_b)$, employing the following reduction formula for the two-body part:

\begin{eqnarray}
\label{2bodyMEx}
\frac{D_2^x}{\sqrt{(2J+1)(2J'+1)}}=
\end{eqnarray}
\begin{eqnarray*}
e_2^\pi(2j_a,2j_c)\left\{
\begin{array}{rrr}
  j_a & j_c &  2    \\
  J' & J & j_d  \\
  \end{array}
\right\}
\langle j_a^\pi || O_1^{\rm bare}(E2) || j_c^\pi \rangle \delta_{b,d}f_{abJ'} +
\end{eqnarray*}
\begin{eqnarray*}
e_2^\nu(2j_b,2j_d)\left\{
\begin{array}{rrr}
  j_b & j_d &  2    \\
  J' & J & j_a  \\
  \end{array} \right\}
\langle j_b^\pi || O_1^{\rm bare}(E2) || j_d^\pi \rangle \delta_{a,c}f_{cdJ}.
\end{eqnarray*}
We determine the optimal values of the secondary effective charges, $e^\rho_2(2j_a,2j_b)$,  by performing
a $\chi^2$ minimization procedure for the following quantity
$$\chi^2=\sum_{j_{a,b,c,d},J,J'}(D_2-D_2^x)^2,$$
where the values of $D_2$ are shown in Table \ref{2body-e2} and the values of $D_2^x$ are
given by Eq.(\ref{2bodyMEx}).
The optimal values of $e^\rho_2(2j_a,2j_b)$ are shown in Table \ref{1body-e2}.
As noted above, by using Eq.(\ref{cha1}) we have calculated the $j$-dependent
one-body effective charges, $e_1^\rho(2j_a,2j_b)$,
 as a function of increasing model-space size, as
shown in Fig. \ref{fig:eff_charges_1body}.

 Similar
 to the case of the one-body E2 matrix elements, we employ the relation (\ref{fit}) to estimate the converged
 values of the secondary effective charges. We take into account the result for $N_{\rm max}$ values
 in a range from 2 to 12. The extrapolated values for $N_{\rm max} =16$ (in italics) are shown in
 Table  \ref{fig:eff_charges_1body}. The estimated converged values are shown in the last 2 columns.

 Note that the neutron secondary effective charges renormalize weakly and tend to an estimated converged value of
 $e_2^\nu(3,3)=0.211$ and $e_2^\nu(3,1)=0.294$. The proton secondary effective charges, on the other
 hand, only slowly converge to the values of $e_2^\pi(3,3)=0.632$ and $e_2^\pi(3,1)=0.567$ and,
 even in the $N_{\rm max}=16$, case constitute only about 50$\%$ of corresponding converged values.

The total effective charges are given by
$e_t^\rho(2j_a,2j_b)=e_1^\rho(2j_a,2j_b)+e_2^\rho(2j_a,2j_b)$.
  In this equation, $e_t^\rho(2j_a,2j_b)$ represents the total effective charge and
   $e_{1(2)}^\rho(2j_a,2j_b)$ refers to the one- (two-) body component of the
total effective charge, respectively.
These results are plotted in Fig. \ref{fig:eff_charges_total_new}.
We see that
the two-body component contributes a small amount to the one-body component,
although it is larger for the proton than for the neutron.
The estimated converged values are shown in the last 2 columns
  in the Table  \ref{fig:eff_charges_1body}.

 The secondary charges enhance the proton one-body effective charges by about 25 $\%$ and the
 neutron one-body effective charges by about 100 $\%$. We see, therefore, that the neutron total
 effective charges are strongly renormalized by the secondary effective charges, when
  compared to the proton effective charges.

\begin{figure}[b]
	\centering
		\includegraphics[scale=0.20]{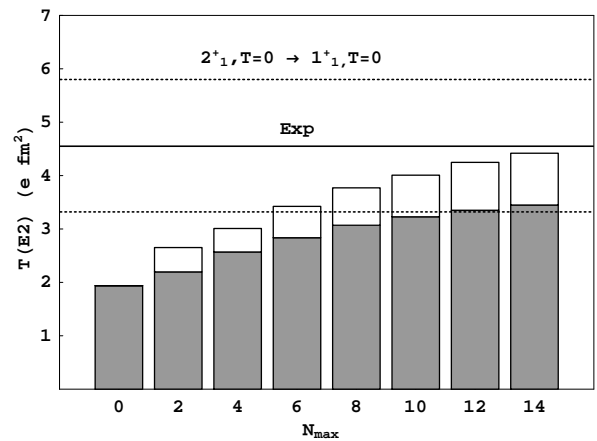}
	\caption{The one- and two-body contributions, as a function of increasing model
space size $N_{\rm max}$, are shown in grey and white, respectively,
for the isoscalar transition $2^+_1(T=0) \rightarrow 1^+_1(T=0)$ matrix element.
}
	\label{fig:E2_211_A6_Nmax012}
\end{figure}


Finally, it should be noted that we have discussed the averaged calculated effective charges. However, particular
 observables may be sensitive to matrix elements, which have been averaged out when calculating secondary
effective charges. In the next subsection we will examine several observables with respect to the two-body
degrees of freedom of the E2 operator.

\subsection{Role of two-body components for $E2$ transitions in $^6$Li}

Here we analyze the nature of the $E2$ transitions in $^6$Li. There are only one-body and two-body contributions
to total $E2$ matrix elements in the case of two valence particles in the p-shell. Thus, we may determine exactly
the one-body and two-body (without using secondary charges) contents of the $E2$ matrix elements.
 In Fig. \ref{fig:E2_211_A6_Nmax012} we show the one- and two-body contributions to the matrix element of
  the isoscalar transition $2^+_1(T=0) \rightarrow 1^+_1(T=0)$ as a function of $N_{\rm max}$.
\begin{figure}[b]
	\centering
		\includegraphics[scale=0.20]{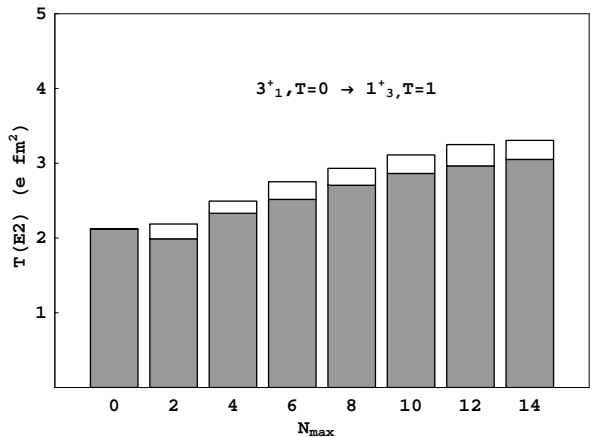}
	\caption{The one- and two-body contributions, as a function of
increasing model space, are shown in grey and white, respectively
 for the isovector transition $3^+_1(T=0) \rightarrow 1^+_3(T=1)$ matrix element.
}
	\label{fig:E2_313_A6_Nmax012}
\end{figure}
 When the model space becomes large, the two-body contribution to this matrix element
  becomes a significant contribution. This indicates that isoscalar transitions are
   renormalized strongly and that higher body correlations are essential in
   accurately calculating the $E2$ operator. This is not the goal of this paper to
 describe the experimental data, but for this particular transition,
 we are able to reproduce the experimental result,  within experimental error,
 provided that the model space is large enough. It is also interesting to estimate a
 converged value of this and other E2 matrix elements using an exponential plus constant
 fit as, we did above for the effective charges.  We compare the results of the fit and
 available experimental data in Table \ref{e2_observ}.
 \begin{table}[tbp]
 \caption{\label{e2_observ} The calculated and experimental E2 reduced matrix elements for $^6$Li.}
 \begin{tabular}{cc|cccc}
 \hline
 \hline
$J_i,T$ & $J_f,T$  & \multicolumn{4}{c}{$\langle J_i ||E2|| J_f \rangle$, (efm$^2$)} \\
\hline
      &         &  \multicolumn{2}{c}{$N_{\rm max}$} & Eq.(\ref{fit})  & Exp. \\
      &         &  12 & 14 & $N_{\rm max} \rightarrow \infty$ &   \\
 \hline
$2^+_1,0$ & $1^+_1,0$   & 4.25  & 4.42 & 5.59 &  4.5(1.3) \\
$3^+_1,0$ & $1^+_1,0$   & 4.96  & 5.16 & 6.80 &  8.6(3) \\
\hline
\hline
\end{tabular}
\end{table}
The results presented in Table III indicate that the extrapolated theory gives a reasonable
estimate of the
available experimental data, taking into account that theoretical NNN forces have not
been included in this study.

In Fig. \ref{fig:E2_313_A6_Nmax012} we show the one- and two-body contributions to the
matrix element of the isovector transition $3^+_1(T=0) \rightarrow 1^+_3(T=1)$
as a function of $N_{\rm max}$. Note that in this case the two-body contribution is relatively
small and that the total
 matrix element renormalizes very weakly as the model space increases. Therefore, we can see an
 indication that isovector transitions are not strongly-dependent on higher-body correlations.
 Such behavior for the isovector transitions has been noted before by the authors in \cite{Nav97}.

\subsection{Role of two-body components for the quadrupole moment of $^6$Li}

\begin{figure}[t]
	\centering
		\includegraphics[scale=0.20]{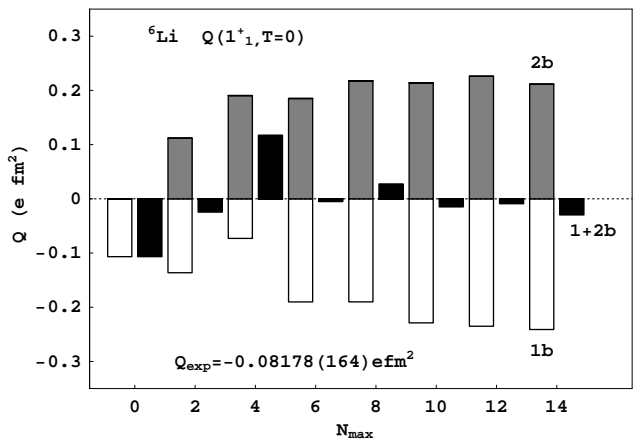}
	\caption{The quadrupole moment of the ground state for $^6$Li ($1^+(T=0)$)
is shown in terms of one- and two-body contributions as a function of increasing model space size.
 The one- and two-body contributions and  total quadrupole moment are depicted as white, grey and black
 histograms, respectively.  The experimentally measured
  quadrupole moment \cite{Cederberg98} is listed on the figure for
  comparison.}
	\label{fig:Q11_A6_Nmax012}
\end{figure}

The quadrupole moment of $^6$Li is notoriously difficult to calculate in the shell-model approach.
 We will now present some insight as to why that may be the case. In Fig. \ref{fig:Q11_A6_Nmax012}
 we show the one- and two-body components of the quadrupole moment as a function of increasing
 model space size. One can immediately draw the interesting conclusion that the one- and two-body
 contributions are similar in size, yet have opposite signs relative to each other.
 The two contributions, thus, tend to cancel each other to give a small resultant quadrupole moment.
 The calculation in the $14 \hbar\Omega$ space yields a quadrupole moment
 $Q_{14\hbar \Omega}[1^+(T=0)]=-0.02971$e fm$^2$. We were also able to calculate the quadrupole moment
 in the larger, $16 \hbar\Omega$ space, where the dimension of the model space exceeds 805 million. The
 obtained  value,  $Q_{16\hbar \Omega}[1^+(T=0)]=-0.02969 \:$efm$^2$, is very close to the one for the
 $14 \hbar\Omega$ space. Note that even in the largest model space (16 $\hbar \Omega$) calculation, our
 calculated quadrupole moment, $Q_{16\hbar \Omega}[1^+(T=0)]=-0.02969 \:$e fm$^2$, is about a factor of
 2.5 times too small, when compared to the experimentally measured value of $Q_{\rm exp}[1^+(T=0)]= -0.0818 \:$efm$^2$ \cite{Cederberg98}.
However, the NCSM calculation with the CD-Bonn potential for $\hbar \Omega =13$ MeV
 in  the $14 \hbar\Omega$ space  results in $Q_{14\hbar \Omega}[1^+(T=0)]=-0.04939$efm$^2$. The NCSM
 results in the same $14 \hbar\Omega$ space with the same frequency, $\hbar \Omega =13$ MeV, but with the
 N3LO interaction \cite{Nav04} yield a slightly different value, -0.06 efm$^2$.
Again, our goal here is not to reproduce experiment, but to understand the physics
of how other physical operators, such as the $E2$ operator, are renormalized due to
truncation of the model space.  In the above case, we conclude that the small quadrupole
moment for $^6$Li comes from an almost total cancellation between the one- and two-body
contributions, arising from the many-body correlations in the
five- and six-body clusters, respectively.


\subsection{Applications to the standard-shell model}

We now turn our attention to testing our effective $E2$ operator represented in terms of
effective charges. In order to do this,
 we perform a SSM calculation of $E2$ transitions for $^7$Li and $^9$Li  and compare the results to
 the NCSM calculations. In the SSM calculations, we use the effective $E2$ operator created from a
 NCSM $6 \hbar \Omega$ calculation specific to the two-body valence cluster approximation for each
  nucleus. We will refer to these effective interactions for $^{7(9)}$Li as the A7(9) interaction,
  respectively.

\begin{figure}[t]
	\centering
		\includegraphics[scale=0.33,angle=-90]{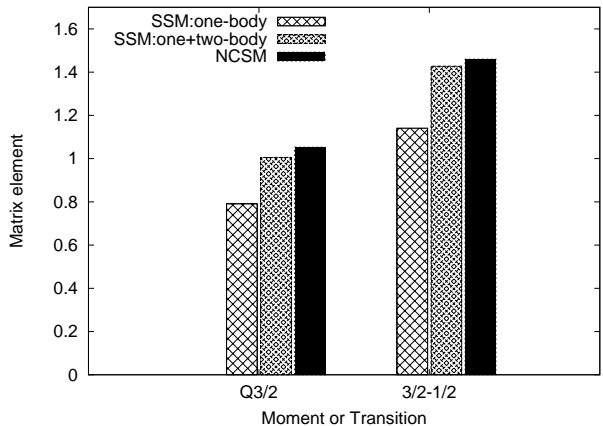}
	\caption{The quadrupole moment (Q3/2) for the $J^\pi=\frac{3}{2}_1^-$ state and
 the $\frac{3}{2}_1^- \rightarrow \frac{1}{2}_1^-$ $E2$ transition matrix element for $^7$Li using
the A7 effective interaction and corresponding effective charges. The left most column
  (criss-crossed filling) refers to a SSM calculation using only the one-body effective
  charges. The middle column (zig-zag filling) refers to a SSM calculation using
   the total effective charges. The right most column (solid black filling) refers to the
   full NCSM calculation for the same transition at $N_{\rm max}=6$.}
	\label{fig:Li7_A7_ssm}
\end{figure}

\begin{figure}[b]
	\centering
		\includegraphics[scale=0.33,angle=-90]{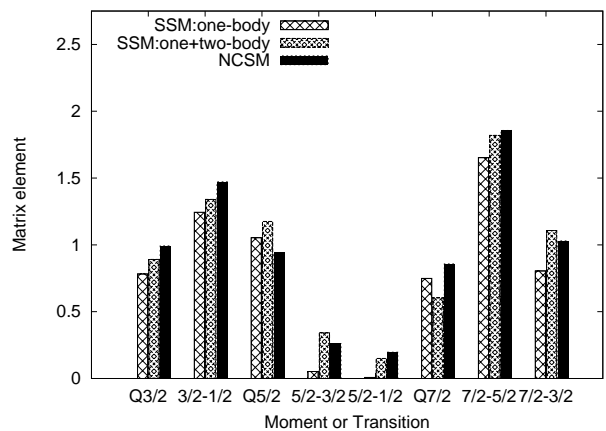}
	\caption{Quadrupole moments and $E2$ transitions for $^9$Li using the A9 effective interaction
and corresponding effective charges.
The quadrupole moments are indicated by a ''QJ'', referring to
the  state with spin $J$. The $E2$ transitions are labeled by values of a spin for initial and final states.
The left most column (criss-crossed filling) refers to a SSM calculation using only the one-body
 effective charges. The middle column (zig-zag filling) refers to a SSM calculation using the total
 effective charges. The right most column (solid black filling) refers to the NCSM calculation for
 the same transition.}
	\label{fig:Li9_A9_ssm}
\end{figure}

The results for $^7$Li and $^9$Li are shown in Figs. \ref{fig:Li7_A7_ssm} and \ref{fig:Li9_A9_ssm}, respectively.
 We can see  in both figures that using only the one-body effective charges is not sufficient
  to reproduce the equivalent NCSM matrix element. The secondary effective charges add
  (in general) a small finite contribution to the SSM matrix element, thereby approximating
  the NCSM matrix element more accurately. This is particularly evident in $^9$Li for the
   $\frac{7}{2}_1^- \rightarrow \frac{5}{2}_1^-$ and
  $\frac{5}{2}_1^- \rightarrow \frac{3}{2}_1^-$ transitions and the quadrupole moment of
   the $\frac{3}{2}_1^-$ state. However, in the case of the quadrupole moment for
   the $J^\pi_n=\frac{5}{2}_1^-$ and $\frac{7}{2}_1^-$ states, for instance, two-body contributions
    modeled using secondary effective charges do not correctly approximate the two-body parts of the E2 effective
     operator. This indicates that exact two-body E2 matrix elements have to be used in order to obtain
   consistent information about interference of one- and two-body contributions.
   Furthermore, there are higher-body correlations, which have been neglected
    ({\em e.g.} the 3-body effective interaction and the 3-body effective E2 matrix elements).

\section{Conclusion}

 We have constructed effective E2 operators for p-shell, which account exactly for up-to 6-body
 correlations in the $14 \hbar \Omega$ space. Our results indicate that 3- and higher-body
 correlations strongly renormalize the E2 operators, enhancing the proton part by about
 70$\%$ and the neutron part by about 40$\%$ relative to the bare proton part of the E2 operator.
 Using 1-body valence cluster (1BVC) and 2-body valence cluster (2BVC) approximations,
 we have decomposed the effective E2 operator into
 one-body and two-body parts, where the effective one-body part accounts for up-to 5-body
 correlations in the $16 \hbar \Omega$ space and the effective two-body for residual
 6-body correlations in the  $14 \hbar \Omega$ space. We have found that the proton two-body part
 of the effective E2 operator constitutes on average about 17$\%$ of the total proton
 renormalization, while the neutron two-body renormalization is about 50$\%$ of the total
 neutron enhancement. Furthermore, we noted that the renormalization for the one-body nondiagonal
 matrix element, $\langle p_{3/2} || E2 || p_{1/2} \rangle$, is about 20$\%$ stronger than
 for the diagonal one, $\langle p_{3/2} ||E2|| p_{3/2} \rangle$.
 We have shown that the two-body part of the effective E2 operator can be
 accounted for reasonably well by introducing secondary effective quadrupole charges.
This approximation can be
 used when the E2 matrix elements are on the scale of ~1.0 efm$^2$. However,
 for small matrix elements ($<$0.2 efm$^2$) the effective two-body part has to be calculated
 exactly to reproduce the considerable two-body effects.
The best illustration of this effect is the case
 of the $^6$Li ground state, when one-body and two-body contributions interfere
 destructively, resulting in a nearly vanishing quadrupole moment.

 We have also shown that our effective E2 operators can be used to approximately
calculate the quadrupole moment and E2 transitions  for heavier nuclei in the SSM
formalism (see for, e.g., Fig.(\ref{fig:Li7_A7_ssm}) and Fig.(\ref{fig:Li9_A9_ssm})).
Such calculations are useful in predicting what a NCSM calculation for the E2 operator
would yield, provided it could be carried out on a sufficiently large computer.
Recall that near the center of a major shell, such as the p-shell, the number of
basis states involved in performing a converged NCSM calculation would be
computationally expensive. However, as we have shown, a SSM calculation is
easily performed and gives a good estimate of what the NCSM result for the E2 operator would be.

 \section{Acknowledgments}We thank the Department of Energy's Institute for Nuclear Theory
 at the University of Washington for its hospitality and the Department of Energy for partial
  support during the completion of this work.
B.R.B. and A.F.L. acknowledge partial support of this
work from NSF grants PHY0244389 and PHY0555396; P.N.
acknowledges support in part by the U.S. DOE/SC/NP
(Work Proposal N. SCW0498) and U.S Department of Energy Grant DE-FC02-07ER41457;
Prepared by LLNL under  Contract No. DE-AC52-07NA27344.
J.P.V. acknowledges support from U.S. Department of Energy
Grants DE-FG02-87ER40371, DE-FC02-07ER41457, and DE-FC02-09ER41582;  B.R.B. thanks the GSI Helmholzzentrum
f\"ur Schwerionenforschung Darmstadt, Germany,
for its hospitality during the preparation of this
manuscript and the Alexander von Humboldt Stiftung
for its support.


\begin{thebibliography}{100}
\bibitem{Nav00} P. Navratil, J.P.Vary, B.R.Barrett, Phys. Rev. Lett. {\bf 84}, 5728 (2000); Phys. Rev. C. {\bf 62}, 054311 (2000).
\bibitem{Nav03} P. Navratil and W.E.Ormand, Phys. Rev. Lett. {\bf 88}, 152502 (2002);
                 Phys. Rev. C. {\bf 68}, 034305 (2003).
\bibitem{Ste05} I. Stetcu, B.R.Barrett, P.Navratil, J.P.Vary,   Phys. Rev. C. {\bf 71}, 044325 (2005).
\bibitem{Nog06} A. Nogga, P.Navratil, B.R.Barrett, J.P.Vary, Phys. Rev. C. {\bf 73}, 064002 (2006).
\bibitem{Nav07} P. Navratil, V. G. Gueorgiev, J.P. Vary, W. E. Ormand, and A. Nogga,  Phys. Rev. Lett. {\bf 99}, 042501 (2007).
\bibitem{Pie02} S. C. Pieper, K. Varga, and R. B. Wiringa, Phys. Rev. C 66, 044310 (2002).
\bibitem{Pie01} S. Pieper and R. B. Wiringa, Annu. Rev. Nucl. Part. Sci. {\bf 51}, 53 (2001).
\bibitem{Kow04} K.Kowalski, D.J.Dean, M.Hjorth-Jensen, T.Papenbrock, P.Piecuch, Phys. Rev. Lett. 92, 132501 (2004).

\bibitem{Rot07} R. Roth and P. Navratil, Phys. Rev. Lett. {\bf 99}, 092501 (2007).

\bibitem{Lis08} A. F. Lisetskiy, B. R. Barrett, M. K. G. Kruse, P.Navratil, I. Stetcu, J. P. Vary, Phys. Rev. C. {\bf 78}, 044302 (2008).
\bibitem{Mac01} R. Machleidt, Phys. Rev. C. {\bf 63}, 024001 (2001).
\bibitem{Cou99} E. Caurier and F. Nowacki, Acta. Phys. Pol. B30, (1999)  705.
\bibitem{Mar99} E. Caurier, G. Martinez-Pinedo, F. Nowacki, A. Poves, J. Retamosa, and A. P. Zuker, Phys. Rev. C {\bf 59}, 2033 (1999).
 \bibitem{Cou01} E. Caurier, P. Navratil, W. E. Ormand, and J.P. Vary, Phys. Rev. C {\bf 64}, 051301(R) (2001).
\bibitem{Bruss77} P. J. Brussaard and P.W.M. Glaudemans, {\em Shell model applications in nuclear spectroscopy}, North-Holland Publishing Company, (1977); p.417.
\bibitem{For08} C. Forss\'en, J. P. Vary, E. Caurier, and P. Navratil, Phys. Rev. C {\bf 77}, 024301 (2008).

\bibitem{Nav97}  P. Navratil, M. Thoresen, and B. R. Barrett, Phys. Rev. C. {\bf 55}, R573 (1997).





\bibitem{Cederberg98} J. Cederberg \emph{et al}, Phys. Rev. A. {\bf 57}, 2539 (1998).
\bibitem{Nav04}  P. Navratil and E. Caurier, Phys. Rev. C {\bf 69}, 014311 (2004).
\end{thebibliography}
\end{document}